# Air-sea interaction described by bilayer networks

Ai-xia Feng    Zhi-qiang Gong    Qi-guang Wang    Guo-lin Feng

*College of Atmospheric Science, Lanzhou University, Lanzhou, 730000, P. R. China*

**Abstract** – We introduce *bilayer networks* in this paper to study the coupled air-sea systems. It proved true that the framework of bilayer networks is powerful for studying the statistical topology structure and dynamics between the fields of ocean and atmosphere. Based on bilayer network, we identify the key correlation regions within and between the air-sea systems. A coupled mechanism between Asia monsoon circulation and Walker circulation is proposed to explain the high correlation phenomenon between the regions on air-sea interaction. We also identify the key regions, which influence the air-sea systems more strongly. The new framework uncovers already known as well other novel features of the air-sea systems and general circulation. It is fruitful to apply the complex networks theory and methodology to understand the complex interactions between the ocean and the atmosphere.

**Introduction.-** The last fewer decades have witnessed the widespread application of complex networks to ecological, social, biological and technological systems [1-3]. Most of the studies focused on representation the networks by one kind of nodes or interaction. However, it began to realize that super-networks or networks of networks could be a more appropriate way to describe the complex systems. The whole network could be considered as the interaction between and within subnetworks. The well-studied networks by the notation are the mammalian cortex networks [4-5] and the infrastructure networks [6-10]. In transportation networks, the authors introduced layered model to describe the system, in which two network layers were used to represent the physical infrastructure and the traffic flows [8-10]. Recently, interconnecting bilayer networks were proposed to study the wide range empirical networks [11]. These are important works for the further development of the complex systems.

The climate networks were also well-studied to reveal the spatio-temporal structure of the climatic variation and the mechanism of the climate dynamics over the globe or a region [12-31]. The works also include the studying the role of teleconnections [15], the impact of the El niño and La niña phenomenon [19-20], the climate shifts [14, 21], and the extreme events [30] of the climate in the perspective of complex networks. Among the works, coupled subnetworks were applied to study vertical topological structure of the geopotential height for understanding the circulation [26]. In this paper bilayer networks were applied to study the air–sea interaction, which is a fascinating, tough and important issues in climate study. In the bilayer networks, the networks are divided into lower layer and upper layer subnetworks. In our research, the surface sea temperature is regarded as the lower layer, and 850hPa geopotential height level as the upper layer. The air-sea systems can be studied by investigating the inner-layer and cross-layer properties to reveal their topological and dynamical structure.

**Data and methodology.-** The data used here are the monthly averaged geopotential height of reanalysis Ⅰ data with a latitude-longitude resolution of 5×5 degrees, and the monthly averaged surface sea temperature (SST) from the NOAA of V3b with latitude-longitude resolution of 4×4 degrees between 60°S and 60°N. The grid points of geopotential height field and SST field are 2664 and 2790 respectively. For each grid point, monthly values are from January 1948 to December 2010. To minimize the bias introduced by the external solar forcing common to all time series, the anomaly time series calculated to remove the annual cycle.

To study the coupled air-sea systems by complex networks, the grid points will be assumed to be the nodes of the network denoted by a set $V = \{v_1, v_2, \mathbf{L}, v_N\}$. There are distinct climatological variables in the networks, therefore we denote the nodes by two layers. One layer is referred to the nodes of the SST field as lower layer by $V_1 = \{i_1, i_2, \mathbf{L}, i_{N_1}\}$, where $i$ and $N_1$ represent the nodes and the total node number of the layer. The value of $N_1$ is equal to 2790. The other layer is referred to the nodes of 850hPa geopotential height field as upper layer by $V_2 = \{j_1, j_2, \mathbf{L}, j_{N_2}\}$, and the symbols are similar to the lower layer. While the value of $N_2$ is equal to 2664. By definition, we have $V_1 \mathbf{U} V_2 = V$ and $N = N_1 + N_2$. In order to introduce the links between nodes, the Pearson correlation coefficient $r$ at non-lag between the anomaly time series of all possible pairs of nodes was calculated. The pair of nodes has an edge if the absolution value of correlation coefficient greater or equal to 0.5. There exist three kind of edges, the edges among the lower layer denoted by set $E_1 = \{e_1(ii), e_2(ii), \mathbf{L}, e_{M_1}(ii)\}$, the edges among the upper layer denoted by set $E_2 = \{e_1(jj), e_2(jj), \mathbf{L}, e_{M_2}(jj)\}$ and the edges connecting the two layer called cross edges denoted by set $E_{12} = \{e_1(ij), e_2(ij), \mathbf{L}, e_{M_{12}}(ij)\}$ (the nodes connecting the cross edges are called cross nodes), then $E_1 \mathbf{U} E_2 \mathbf{U} E_{12} = E$ and $M = M_1 + M_2 + M_{12}$. The density of the edges is defined as $r = \frac{2|M|}{N(N-1)}$ for the whole network, thus $r_1 = \frac{2|M_1|}{N_1(N_1-1)}$ for the lower layer edges, $r_2 = \frac{2|M_2|}{N_2(N_2-1)}$ for the upper layer edges and $r_{12} = \frac{2|M_{12}|}{N_1 N_2}$ for the cross edges to measure the proportion that the nodes could be connected.

The node degree and cluster coefficient are popular physical quantity to describe the topology properties of the nodes. For the bilayer networks, we focus on the properties of the lower layer, the upper layer and the cross properties between the two layers. The node degree of lower layer, the upper layer and the cross node degree between the two layers are defined as:

$$k_i^1 = \sum_{l \in V_1, l=1}^{l=N_1} a_{il}, \quad k_j^2 = \sum_{l \in V_2, l=1}^{l=N_2} a_{jl},$$

$$k_i^{12} = \sum_{l \in V_2, l=1}^{l=N_2} a_{il}, \quad k_j^{12} = \sum_{l \in V_1, l=1}^{l=N_1} a_{jl}. \quad (1)$$

While the weighted degrees are:

$$w_i^1 = \frac{\sum_{l \in V_1, l=1}^{l=N_1} a_{il} \cos f_l}{\sum_{l \in V_1, l=1}^{l=N_1} \cos f_l}, \quad w_j^2 = \frac{\sum_{l \in V_2, l=1}^{l=N_2} a_{jl} \cos f_l}{\sum_{l \in V_2, l=1}^{l=N_2} \cos f_l},$$

$$w_i^{12} = \frac{\sum_{l \in V_2, l=1}^{l=N_2} a_{il} \cos f_l}{\sum_{l \in V_2, l=1}^{l=N_2} \cos f_l}, \quad w_j^{12} = \frac{\sum_{l \in V_1, l=1}^{l=N_1} a_{jl} \cos f_l}{\sum_{l \in V_1, l=1}^{l=N_1} \cos f_l}. \quad (2)$$

Where $a_{il}$ or $a_{jl}$ is the element of the adjacent matrix with $a_{il}$ or $a_{jl} = 1$ if there exists edge between $i$ and $l$ or $j$ and $l$, otherwise $a_{il} = 0$ or $a_{jl} = 0$. The node degree measures the centrality of the node in the network, and the weighted degree is the same but minimizing the bias induced by different grids representing different areas on the earth. The cluster coefficients for them are as follow:

$$C_i^1 = \frac{1}{k_i^1(k_i^1-1)} \sum_{p \neq q \in V_1} a_{ip} a_{pq} a_{qi},$$

$$C_j^2 = \frac{1}{k_j^2(k_j^2-1)} \sum_{p \neq q \in V_2} a_{jp} a_{pq} a_{qj},$$

$$C_i^{12} = \frac{1}{k_i^{12}(k_i^{12}-1)} \sum_{p \neq q \in V_2} a_{ip} a_{pq} a_{qi},$$

$$C_j^{12} = \frac{1}{k_j^{12}(k_j^{12}-1)} \sum_{p \neq q \in V_1} a_{jp} a_{pq} a_{qj}. \quad (3)$$

If the node degree is equal to 0 or 1, then the cluster coefficient is zero. It provides the probability linking of the nearest neighbors of a node, and indicates the closeness of a structure (community in society).

**Results and discussion of the air-sea bilayer climate networks.-** The densities of the bilayer climate networks constructed by us are summarized in Table 1. The densities of the upper layer and the lower layer are pretty the same, but the cross density is much smaller about twentieth of that of the two separate layer. That is to say, there is physical separation of the underlying dynamics of the SST field and 850hPa geopotential height field, and the global air-sea interaction centers at some key regions. That is why we introduce bilayers networks to study the air-sea systems.

Table 1: The linking probability of the air-sea coupled climate networks: the lower layer refers to the SST field, and the upper layer refers to the 850hPa geopotential height field.

| networks | whole | lower layer | upper layer | cross |
|---|---|---|---|---|
| density | 0.024 | 0.038 | 0.053 | 0.002 |

The weighted node degree of the lower layer is color contoured in Fig.1. The spatial heterogeneity of the averaged linear correlation structure in SST field is clearly distinct. The weighted node degree is significantly larger in the tropical Indian Ocean and Mid-east Pacific Ocean than other ocean regions. They are key regions affecting the climatic systems. In order to identify the relationship between them, the SST anomaly evolution of the two oceans is illustrated by Fig.2 (only the periods from 1971 to 2010 are displayed for clearness). It shows the positive correlation of SST variations in the two ocean regions. The cross correlation coefficient of them is 0.5566 from 1948 to 2010. That is rather larger than the value of statistically significant above the 99% level. That uncovers that the changes of the two oceans occur synchronously. It is hard to be explained by the inner-process of the Oceans. The phenomenon will be explained later in the paper. The structure of weight node degree centrality in the upper layer is shown in Fig.3. Similar to the SST field the 850hPa geopotential height field also has a stronger correlation in tropical regions than the mid-high latitudes.

That implies the centrality role of the equatorial regions. The strongest correlation of the 850hPa is in the region of the middle Africa, which may contribute to the Somali jet what makes the region closely correlated with the tropical Indian Ocean and the India Peninsula.

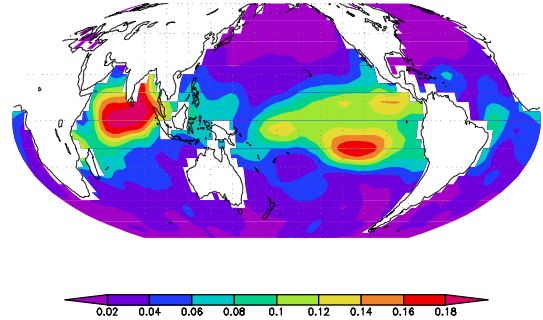

Fig. 1: The geographic distribution of the weighted node degree in the lower layer subnetwork.

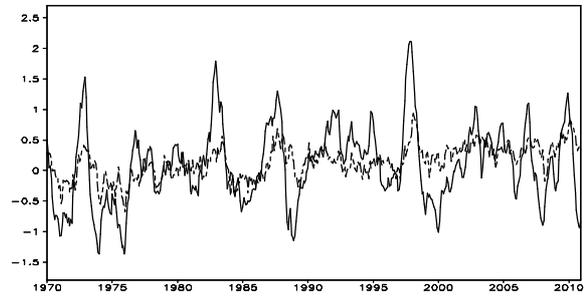

Fig. 2: The SST anomaly in the Pacific Ocean (-8°S-8°N, 180-272°E, the solid line) and the Indian Ocean (-8°S-8°N, 44°E -100°E, the dashed line) from 1971 to 2010.

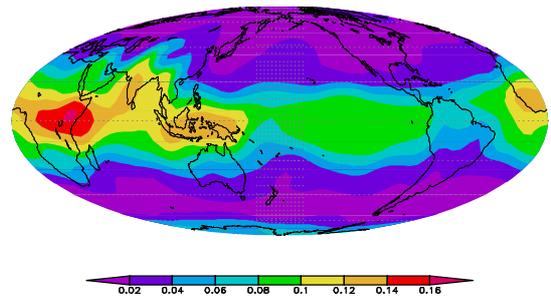

Fig. 3: The same as Fig.1 but in the upper layer subnetwork.

The weighted cross node degree of the lower layer and upper layer are illustrated in Fig.4 and Fig.5. The regions of largest weighted cross node degree locate in the tropical mid-east Pacific Ocean in the SST field and islands isolation the tropical Indian Ocean and Pacific Ocean in the 850hPa circulation field, which imply the important role of the regions in the air-sea interaction process. The graph of the strong interaction of the bilayer networks is just as fig.6 shown. It visualized shows the inner-layer and cross layer relationships of the air-sea interactions. In the lower layer, the high correlation regions are between the tropical Indian Ocean and mid-east Pacific Ocean. Other ocean regions correlated to the two oceans are the tropical Atlantic Ocean, the west Pacific Ocean, and the mid-high latitudes of south Indian Ocean and Pacific Ocean. The pattern of the SST field inner-correlation is like an irregular double stars and the central discs are the two tropical oceans. The shape of interactions for the 850hPa likes a band center in the African continent, the tropical Indian Ocean, south of Eurasia, equatorial west Pacific and the tropical Atlantic. For the cross layer interaction, the main interactions are between the tropical islands (Malayan Peninsula and Indonesia archipelago) area and Indian Ocean area, and between the tropical islands area and Pacific Ocean area. Furthermore, the later interaction is much closely. The correlated coefficients between the anomaly geopotential height of the islands area (12°S-8°N，115°E-150°E, in 850hPa) and the anomaly SST of the two oceans are 0.6051 and 0.6889 respectively.

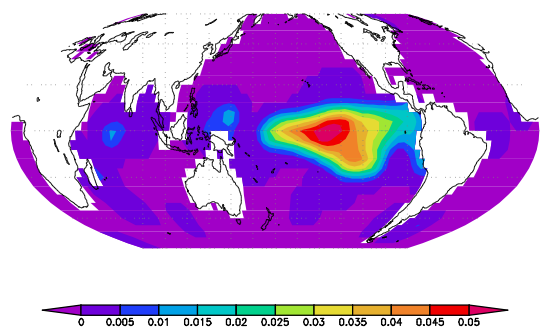

Fig. 4: The geographic distribution of the weighted cross node degree in the SST field.

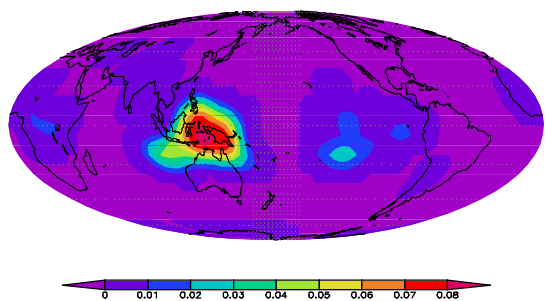

Fig. 5: The same as Fig.4 but in the upper layer.

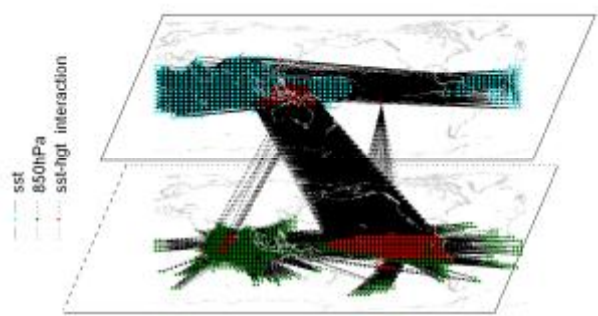

Fig. 6: The graph of bilayer air-sea interaction networks: The dots with olive color and cyan color represent the nodes with weight node degree greater than 0.18 for the lower layer and 0.14 for the upper layer. The red dots represent the cross nodes with weight node degree greater than 0.06. The black dash or solid lines

represent edges.

The high non-lag positive correlations between the two oceans regions in the SST field and the tropical islands in 850hPa reveal the underlying dynamical mechanism between them. The possible mechanism is that the SST changes in the two oceans and the geopotential height variations over the tropical islands are connections by the atmosphere circulation. The circulation cells over the oceans are the Asia monsoon circulation (over the Indian Ocean) and walker circulation (over the Pacific Ocean). The anomaly changes with time of the two circulation cells are shown in Fig.7. They are negative correlated and the coefficient is -0.3098, and the value is also higher than that of statistically significant above the 99% level. Therefore, the variations of the two circulation cells are synchronous, but in direction they are contrary. The process of air-sea interaction in the tropical Indian-Pacific ocean regions can be described by the Fig.8 and Fig.9. When the SST anomaly of two oceans is positive, the anomaly of the islands' geopotential height is positive as well which implicates the surface pressure is relatively high in the island and relatively low in the oceans. So the anomaly flows are upward over the oceans, but when they arrives to the 850hPa with the u wind of Asia monsoon and Walker circulation then they change to be level wind flowing towards the region over the islands, after that they subsides in the region. Because of the higher pressure in the region, the surface flows are blowed from the islands to the two Oceans because of the higher pressure in the region and forms the closure atmosphere circulation. The Asia monsoon circulation is clockwise and the Walker circulation counter-clockwise just as Fig.8 shown. When the anomaly is negative, the situation is opposite just as Fig.9 illustrated.

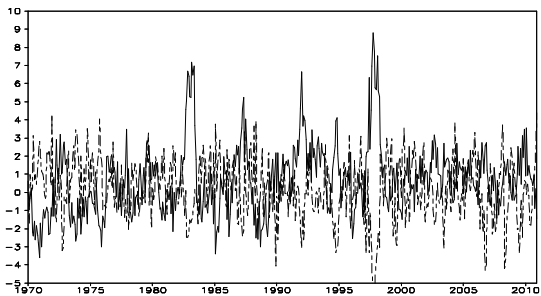

Fig. 7: The same as Fig.2 but for the anomaly u wind of 850hPa.

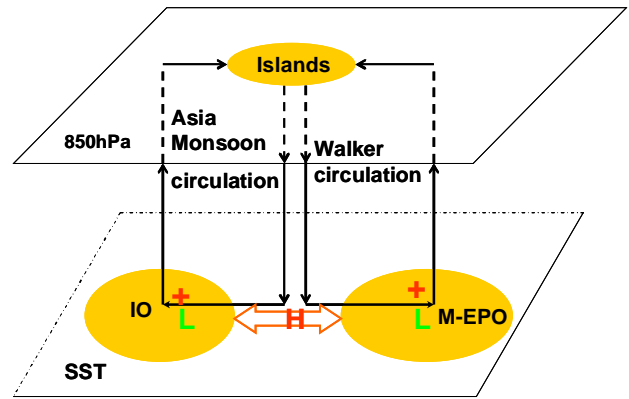

Fig. 8: Schematic of the air-sea interaction mechanism between the Indian Ocean (IO), the Mid-east Pacific (M-EPO) and the atmosphere over the islands isolation them when their anomalies are positive.

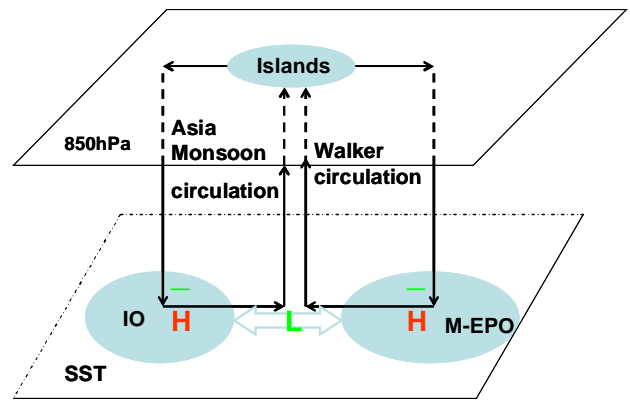

Fig. 9: The same as Fig.8 but corresponding to anomalies are negative.

The mechanism of the tropical air-sea interaction is in keep with the "GIP" model very well, which has successfully explained significant positive correlation (without lag) in SST anomalies between the equatorial Indian Ocean and the eastern equatorial Pacific Ocean [32-33]. It attributed the sound no-delay correlation between the two tropical oceans to the strong coupling between the monsoonal zonal circulation over the equatorial Indian

Ocean and the Walker circulation over the Pacific Ocean. The model proposed the two coupled circulations works in a way like a pair of gear operating over the equatorial Indian and Pacific and revealed the "gearing point" locating by the Indonesia archipelago, which correspond with our cross interactions by complex networks and the air-sea interaction mechanism. Furthermore, our method reveals that the air-sea interaction in the Pacific region is much stronger than the Indian region.

In order to know the closeness of the interaction between the nearest neighbors, the cluster coefficient distribution is calculated as well. The largest inner-cluster coefficient of SST field (Fig.10) locates at the tropical Pacific Ocean (rather narrow), the west coastline of South America and some east coastline of Africa indicating a compact inner-interaction. The smallest inner-cluster coefficient of SST field is in the tropical Indian Ocean, which indicate a loose interactions between the regions correlated to the Ocean. Other two smaller cluster regions are the north-east Pacific and south of Pacific marked by red cycles. The SST anomaly in the regions with larger cluster could have greater impact on the SST field than those with smaller cluster, because the former is related closely to other regions. The inner-cluster coefficient distribution in 850hPa geopotential height field is illustrated by Fig.11. The largest cluster coefficients locate in the Antarctic continent. Other larger regions are over the mid-east Pacific Ocean, Indian Ocean and some regions scattered in the Atlantic Ocean, the middle of South America, the south of Africa and the Eurasia. The interaction structure of these regions, playing a great role in the circulation, is more compacter and more stable. The cross-cluster coefficient distribution of SST field is shown in Fig.12. The large cross-cluster coefficient regions mainly locate at the tropical Indian Ocean and Pacific Ocean, which reveals these regions influence the upper atmosphere greatly. The larger cross-cluster coefficients in upper layer are over the regions of mid-east Pacific and the islands isolating the tropical Indian Ocean and Pacific Ocean. They are key regions in the air-sea interaction systems.

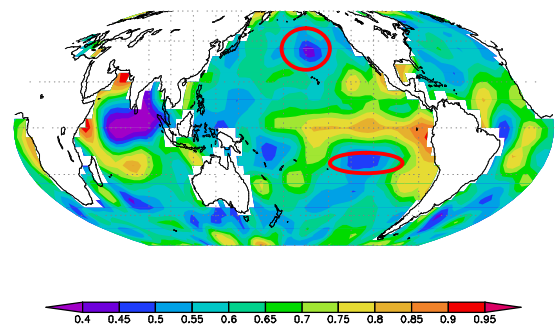

Fig. 10: The cluster distribution of the lower layer.

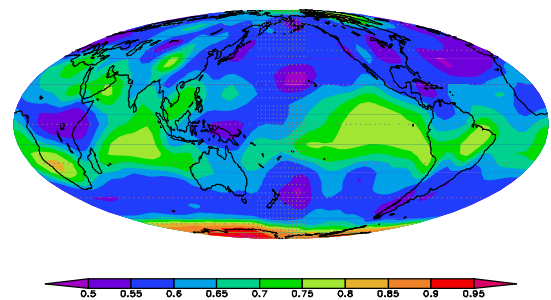

Fig. 11: The cluster distribution of the upper layer.

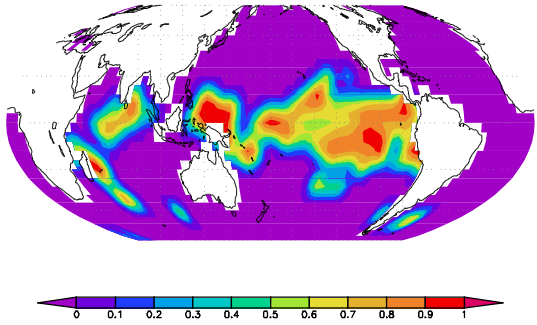

Fig.12: The cross cluster distribution of the lower layer.

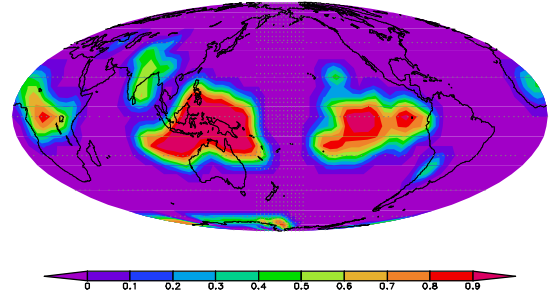

Fig.13: The cross cluster distribution of the upper layer

**Conclusions.-** In summary, we introduced bilayer networks to study the air-sea interaction systems and investigated the properties of the coupled climate networks. The technique focused on revealing the topological structure and dynamical mechanism of the air-sea systems. The weighted node degree of the bilayer networks revealed the correlation structure in the SST field, the 850hPa geopotential height field and the interaction between them. We found the centrality regions of the correlation locate at tropical Indian Ocean and Pacific Ocean in the SST field, and tropics in the 850hpa field. The cross interaction regions of the two fields were identified in the two Ocean in the SST field and the areas over the islands isolating the two Ocean in the 850hPa field. A mechanism was proposed to explain the phenomenon and connect it to the coupled Asia monsoon circulation and Walker circulation, which was corresponding with the "GIP" model well. Meanwhile our method uncovered that the air-sea interaction in the Pacific region was much stronger than the Indian region.

The cluster coefficient provided the closeness of correlated neighboring nodes of a node. It was found that the low closeness of tropical Indian Ocean' correlated neighbors, but that for the tropical Pacific Ocean is high. For the 850hPa field, the high closeness regions of correlated neighbors were mainly lying in the tropical regions and Antarctic continent. The main regions with high crossed-neighborhood closeness locate in the tropical Indian and Pacific Ocean in the lower layer, and in tropical mid-east Pacific Ocean, the tropical islands in (in the middle of the Indian and Pacific Ocean) and their adjacent oceans in the upper layer. Those regions with high cluster coefficient are the key regions in the air-sea systems, because they have closely correlated neighbors who can response the anomaly quickly. The facts prove that it is fruitful to apply bilayer networks to study the air-sea systems.

∗∗∗

We give our thanks to Professor D. R. He and H. X. Cao for constructive suggestion and correction of the paper. We acknowledge Ph.D J. F. Donges for providing useful information. This work is supported by the Chinese National Natural Science Foundation under grant numbers 40930952 and 40875040.